\documentclass[aps,prb,twocolumn,superscriptaddress,showpacs,floatfix,longbibliography]{revtex4-2}
\usepackage{amsmath,amssymb,amsfonts,float,graphics,epsfig,epstopdf,color,verbatim,tabularx,bm,multirow,appendix,hyperref}
\usepackage[normalem]{ulem}
\usepackage{lmodern}
\usepackage{ytableau}
\usepackage{pifont}
\usepackage{color}
\usepackage{bm}
\DeclareMathOperator{\Tr}{Tr}

\begin{document}
	
\title{An integral algorithm of exponential observables for interacting fermions in quantum Monte Carlo simulation}

\author{Xu Zhang}
\affiliation{Department of Physics and HKU-UCAS Joint Institute of Theoretical and Computational Physics, The University of Hong Kong, Pokfulam Road, Hong Kong SAR, China}
\author{Gaopei Pan}
\affiliation{Department of Physics and HKU-UCAS Joint Institute of Theoretical and Computational Physics, The University of Hong Kong, Pokfulam Road, Hong Kong SAR, China}
\author{Bin-Bin Chen}
\affiliation{Department of Physics and HKU-UCAS Joint Institute of Theoretical and Computational Physics, The University of Hong Kong, Pokfulam Road, Hong Kong SAR, China}
\author{Kai Sun}
\email{sunkai@umich.edu}
\affiliation{Department of Physics, University of Michigan, Ann Arbor, MI 48109, USA}
\author{Zi Yang Meng}
\email{zymeng@hku.hk}
\affiliation{Department of Physics and HKU-UCAS Joint Institute of Theoretical and Computational Physics, The University of Hong Kong, Pokfulam Road, Hong Kong SAR, China}

\begin{abstract}

Exponential observables, formulated as $\log \langle e^{\hat{X}}\rangle$ where $\hat{X}$ is an extensive quantity, play a critical role in study of quantum many-body systems, examples of which include the free-energy and entanglement entropy. Given that $e^{X}$ becomes exponentially large (or small) in the thermodynamic limit, accurately computing the expectation value of this exponential quantity presents a significant challenge. In this Letter, we propose a comprehensive algorithm for quantifying these observables in interacting fermion systems, utilizing the determinant quantum Monte Carlo (DQMC) method. We have applied this algorithm to 2D square lattice half-filled Hubbard model and $\pi$-flux t-V model. In Hubbard model case at the strong coupling limit, our method showcases a significant accuracy improvement on free energy compared to conventional methods that are derived from the internal energy, and in t-V model we indicates the free energy offer precise determination of second-order phase transition. We also illustrate that this approach delivers highly efficient and precise measurements of the nth R\'enyi entanglement entropy. Even more noteworthy is that this improvement comes without incurring increases in computational complexity. This algorithm effectively suppresses exponential fluctuations and can be easily generalized to other models.
	
\end{abstract}
\date{\today}
\maketitle

\noindent{\textcolor{blue}{\it Introduction.}---} Exponential observables such as the entanglement metrics and free energy hold a crucial role in unveiling the fundamental organizing principles of strongly correlated systems, as they offer the full access to the partition function and universal conformal field theory (CFT) data of (quantum) many-body systems which are otherwise hard to obtain. Such understandings have been extensively put forward in previous works~\cite{CARDY1988377,srednicki1993entropy,Solfanelli:2023vav,Pasquale,fradkin2006entanglement,CASINI2007183,grover,gioev2006entanglement,swingle2010entanglement,leschke2014scaling,LAFLORENCIE20161,assaad2014entanglement,broeckerRenyi2014,broeckerEntanglement2016,broeckerNumerical2016,albaOut2017,dEmidioEntanglement2020,zhaoScaling2022,zhaoMeasuring2022,d2022universal,chenTopological2022,pan2023stable,da2023controllable,da2023teaching,kitaev2006topological,levin2006detecting,jiangMany2023,liuFermion2023,songDeconfined2023,songResummation2023,liuDisorder2023,zhouIncremental2024,liaoExtract2024,Onsager1944,Schultz1964}. Among these witnesses, entanglement entropy (EE) stands out as a vital metric for probing the behavior of interacting fermion systems in spatial dimension $D\ge 2$~\cite{grover,gioev2006entanglement,swingle2010entanglement,leschke2014scaling,LAFLORENCIE20161,assaad2014entanglement,broeckerRenyi2014,broeckerEntanglement2016,broeckerNumerical2016,d2022universal,pan2023stable,da2023controllable,da2023teaching}. In the case of free Fermi surfaces, the scaling form of EE is well-established through the Widom-Sobolev formula, expressed as $L^{D-1} \log L$, where $L$ represents the linear system size~\cite{gioev2006entanglement,swingle2010entanglement,leschke2014scaling}. However, for interacting fermion systems, the precise scaling form remains elusive, and it is widely held that uncovering this scaling behavior could offer valuable insights into the fundamental CFT data associated with fermionic quantum critical points, low-energy collective modes, and topological information~\cite{fradkin2006entanglement,CASINI2007183,grover,gioev2006entanglement,swingle2010entanglement,leschke2014scaling,LAFLORENCIE20161,assaad2014entanglement,d2022universal,da2023teaching,kitaev2006topological,levin2006detecting,liuDisorder2023,shaoEntanglement2015,mishmashEntanglement2016}. 

As another example of exponential observable, the free energy is a key physics quantity that directly dictates the finite-temperature phase diagram of a many-body system~\cite{Onsager1944,Schultz1964}. For quantum Monte Carlo (QMC) simulations, the free-energy also plays an important role in sign problem, especially in sign bound theory~\cite{loh1990sign,panSign2024,zhangsign2022,zhangpolynomial2023,mondainiUniversality2023,mouBilayer2022,mondainiQuantum2022,taratDeconvolving2022}. Since the lower bound of the average sign amounts to a function of the free energy difference between the original system and the reference system~\cite{zhangsign2022}, calculating free energy of the system helps to understand how the average sign is affected by the free energy of the two systems, for example, to determine the relationship between the average sign and the phase transition~\cite{yan2023universal,mondainiUniversality2023,mouBilayer2022,mondainiQuantum2022,taratDeconvolving2022}.


However, the computation of EE and free energy in 2D interacting fermion systems poses a formidable challenge. The computation hinges on accessing the observables exponentially proportional to $L$ in the many-body partition functions as detailed in prior studies~\cite{Pasquale,assaad2014entanglement,broeckerRenyi2014,broeckerEntanglement2016,broeckerNumerical2016,d2022universal,da2023teaching,pan2023stable,da2023controllable}. To controllably compute these exponential observables within polynomial computational complexity, the determinant quantum Monte Carlo (DQMC) method has emerged as a promising solution~\cite{grover,assaad2014entanglement,drut2015hybrid,drut2016entanglement,d2022universal,da2023teaching,pan2023stable}. Following Grover's pioneering work~\cite{grover} where many-body reduced density matrix is expanded according to the auxiliary field configuration subjected to the fermion Green's function, the algorithmic development of fermion EE has taken a long detour over the years. 

Early attempts didn't carry out the proper important sampling and therefore suffered from the poorly controlled data quality and less optimized computational complexity of $O(\beta N^4)$ where $\beta =1/T$ the inverse temperature and the $N=L^{D}$ the system size~\cite{assaad2014entanglement,broeckerRenyi2014,broeckerEntanglement2016,broeckerNumerical2016}. In Ref.~\cite{drut2015hybrid,drut2016entanglement}, the integral idea has been proposed while the computational complexity is still $O(\beta N^4)$. The recent developments, in particular the incremental algorithm of EE computation~\cite{d2022universal}, manage to reduce the complexity to $O(\beta N^3)$ (the same as general DQMC update) but add the extra procedure of sampling the entanglement area~\cite{da2023teaching,pan2023stable}. In Refs.~\cite{da2023controllable}, sampling of the entanglement area is replaced by incrementally raising the power of Grover's matrix which makes the operation much easier (similar development appeared recently in quantum spin systems~\cite{zhouIncremental2024}), and such polynomial form suppresses the exponential fluctuation of the observable. But there appears no generic guidance on how often the entanglement area should be sampled or how small this polynomial power should be adequate to suppress the exponential variance before actual computing. Therefore, the protocol still appears as {\it{ad hoc}}.

In this Letter, we provide an elegant solution to this challenge of computing exponential observables combining the integral idea and the fast update routine. The integral idea~\cite{drut2016entanglement} is that via simply introducing an auxiliary integral, measurement of exponential observables can be converted to a conventional observable without involving the expectation of any exponentially large (or small) quantities
 	\begin{equation}
 	\begin{aligned}
 		\log \langle  e^{\hat{X}} \rangle = \int_0^1 \mathrm{d}t \langle \hat{X}\rangle_t
 	\end{aligned}
 	\label{eq:eq0}
 	\end{equation}
Here $\langle \ldots \rangle_t$ represents a modified average where the distribution probability is determined by an extra $t$ with the factor of $e^{t \hat{X}}$. By applying this formula, all complications and challenges in measuring exponential observables have been fully eliminated, and the computational complexity becomes identical to the measurement of regular physics observables. Departing from this formula, we introduce a fast update integral algorithm in DQMC and use 2D square lattice half-filled Hubbard model and $\pi$-flux t-V model to illustrate the superior performance and reduced computational complexity of our algorithm. In Hubbard model case, free energy, the 2nd and 3rd R\'enyi EE are measured stably at finite temperature and their values at the low temperature limit match well with the results from density matrix renormalization group (DMRG). In t-V model case, our free energy computation is precise for indicating second-order phase transition.

We foresee this method contributes to the advancement on various directions, including discovering the unknown scaling form of EE for interacting Fermi surface~\cite{jiangMany2023,shaoEntanglement2015,mishmashEntanglement2016}, the CFT information for various fermion deconfined quantum critical points (DQCPs) and symmetric mass generation (SMG), helping to identify fermion phase transitions beyond Landau-Ginzburg-Wilson paradigm~\cite{liuFermion2023,liuDisorder2023}, indicating first-order or second-order phase transition directly from free energy, computing domain-wall free energy in spin glass systems~\cite{Hukushima1999,Hukushima1996}, etc.

\noindent{\textcolor{blue}{\it Formula for exponential observables.}---}
We review the formula for computing the 2nd R\'enyi EE first, and then give similar formula for computing the nth R\'enyi EE and free energy. The details of fast update integral algorithm is given in the next section. From Ref.~\cite{grover}, we know the 2nd R\'enyi EE $S_{A}^{(2)}\equiv-\log(\Tr(\rho_{A}^2))$ (Here $A$ labels set of sites whose degree of freedom is not traced in reduced density matrix $\rho_A$) can be expressed according to auxiliary field configuration $\{s_1,s_2\}$ as
\begin{equation}
S_{A}^{(2)}=-\log \left(\frac{\sum_{s_1,s_2}P_{s_1}P_{s_2}\Tr \left(\rho_{A;s_1}\rho_{A;s_2}\right)}{\sum_{s_1,s_2}P_{s_1}P_{s_2}}\right),
\label{eq:eq1}
\end{equation}
where $P_{s_i}$ is the importance sampling Monte Carlo weight for configuration $s_i$, $\rho_{A;s_i}=\det(G_{A;s_i})e^{c^\dagger \log(G_{A;s_i}^{-1}-I) c}$ is the reduced density matrix and $\Tr(\rho_{A;s_1}\rho_{A;s_2})=\det((I-G_{A;s_1})(I-G_{A;s_2})+G_{A;s_1}G_{A;s_2})$ is the determinant of the Grover matrix~\cite{grover,d2022universal,pan2023stable,da2023controllable}. Here and below, we define the fermion Green's function as $G_{ij}\equiv\left\langle c_i c_j^\dagger \right\rangle$, where $\langle \ldots \rangle$ without subscript always indicates grand-canonical ensemble average which is used in DQMC simulation. Since the EE is generally an extensive quantity dominated by the area law, a direct simulation by using $P_{s_1}P_{s_2}$ as sampling weight and $\Tr(\rho_{A;s_1}\rho_{A;s_2})$ as observable according to Eq.~\eqref{eq:eq1} will certainly give exponentially small values as $e^{-S_{A}^{(2)}} \sim e^{-a l_A}$ where $l_A$ is the boundary length of the entanglement region defined by $A$ set. This is why the direct simulation according to Eq.~\eqref{eq:eq1} is found to be unstable~\cite{assaad2014entanglement,broeckerRenyi2014,broeckerEntanglement2016,broeckerNumerical2016,pan2023stable}, i.e. one is sampling exponentially small observable which can have exponentially large relative variances/fluctuations.

We notice Eq.~\eqref{eq:eq1} can be rewritten in an integral form
\begin{eqnarray}
&&S_{A}^{(2)}=-\log(f(1))+\log(f(0)) \nonumber\\
&&=-\int_{0}^{1} dt \frac{\sum P_{s_1}P_{s_2}\Tr(\rho_{A;s_1}\rho_{A;s_2})^{t} \log(\Tr(\rho_{A;s_1}\rho_{A;s_2}))}{\sum P_{s_1}P_{s_2}\Tr(\rho_{A;s_1}\rho_{A;s_2})^{t}} \nonumber\\
\label{eq:eq2}
\end{eqnarray}
with $f(t) \equiv \sum P_{s_1}P_{s_2}\Tr(\rho_{A;s_1}\rho_{A;s_2})^{t}$ the entanglement partition function. 
Here and below, we omit auxiliary field labels in $\sum$ for notational simplicity. One can see at each $t$, the original observable $\Tr \left(\rho_{A;s_1}\rho_{A;s_2}\right)$ becomes logarithmic so that exponential fluctuations disappear naturally. In fact, one can choose other $f(t)$ satisfying a general integral formula
\begin{equation}
S_{A}^{(2)}=-\int_{0}^{1} dt \frac{\partial \log( f(t))}{\partial t}
\label{eq:eq3}
\end{equation}
where the only requirements for $f(t)$ are $f(1)=\sum P_{s_1}P_{s_2}\Tr(\rho_{A;s_1}\rho_{A;s_2})$ and $f(0)=\sum P_{s_1}P_{s_2}$ (when taking $f(t)=\langle  e^{t\hat X}\rangle $ we recover Eq.~\eqref{eq:eq0}).

Eq.~\eqref{eq:eq2} can also be derived from taking small polynomial power limit of the incremental algorithm~\cite{da2023controllable,da2023teaching,pan2023stable} (see Appendix B for detail). Besides the 2nd R\'enyi EE, any exponentially small (or large) observable for interacting fermion systems in DQMC can be computed in a similar way. As examples, we introduce the formulas for the nth R\'enyi EE $S_{A}^{(n)}$
\begin{eqnarray}
S_{A}^{(n)}&\equiv&-\frac{1}{n-1}\log(\Tr(\rho_{A}^n)) \nonumber\\
&=&\frac{1}{1-n}\int_{0}^{1} dt \frac{\sum P_{s}^n\Tr(\rho_{A;s}^n)^{t} \log(\Tr(\rho_{A;s}^n))}{\sum P_{s}^n\Tr(\rho_{A;s}^n)^{t}} \nonumber\\
&\equiv&\frac{1}{1-n}\left\langle \log(\Tr(\rho_{A;s}^n))\right\rangle_{s_1,s_2,...,s_n;t}, \nonumber\\
\end{eqnarray}
and free energy $F$
\begin{eqnarray}
F&\equiv&-\frac{1}{\beta}\log(Z) \nonumber\\
&=&-\frac{1}{\beta}\int_{0}^{1} dt \frac{\sum W_{s} P_{s}^{t} \log(P_{s})}{\sum W_{s} P_{s}^{t}} \nonumber\\
&\equiv&-\frac{1}{\beta}\left\langle \log(P_{s}) \right\rangle_{s,t}. 
\label{eq:eq6}
\end{eqnarray}
Here $P_{s}^n\equiv\prod_{i=1}^n P_{s_i}$, $\rho_{A;s}^n\equiv\prod_{i=1}^n\rho_{A;s_i}$, $W_s$ is the decouple coefficient independent with Hamiltonian and $P_s$ is the determinant contribution from Hamiltonian (e.g., if we decouple the Hubbard U term like $ e^{\alpha \hat{O}^2}=\frac{1}{4} \sum_{l= \pm 1, \pm 2} \gamma(l) e^{\sqrt{\alpha} \eta(l) \hat{o}}+$ $O\left(\alpha^4\right)$, where $\gamma( \pm 1)=1+\frac{\sqrt{6}}{3}, \gamma( \pm 2)=1-\frac{\sqrt{6}}{3}, \eta( \pm 1)= \pm \sqrt{2(3-\sqrt{6})}$ and $\eta( \pm 2)= \pm \sqrt{2(3+\sqrt{6})}$, then the part of $\frac{\gamma(l)}{4}$ is the $W_s$ term, and  after tracing out the fermion degree of freedom, the determinant part is the $P_s$ term~\cite{assaadevertz2008}).

\noindent{\textcolor{blue}{\it DQMC integral algorithm.}---}
Next we describe how to implement our integral algorithm in the DQMC simulation. To compute the nth R\'enyi EE, one still needs a generic fast update scheme to guarantee the overall $O(\beta N^3)$ computational complexity~\cite{d2022universal}. Firstly let's review the fast update formulae for Green's functions $G \to G'$,
\begin{eqnarray}
	G'(\tau,\tau)&\equiv&G(\tau,\tau)-\frac{1}{R_0} G(\tau,\tau)\Delta(i,\tau)(I-G(\tau,\tau)), \nonumber\\
	G'(\tau,0)&\equiv&G(\tau,0)+\frac{1}{R_0} G(\tau,\tau)\Delta(i,\tau)G(\tau,0), \nonumber\\
	G'(0,\tau)&\equiv&G(0,\tau)-\frac{1}{R_0} G(0,\tau)\Delta(i,\tau)(I-G(\tau,\tau)), \nonumber\\
	G'(0,0)&\equiv&G(0,0)+\frac{1}{R_0} G(0,\tau)\Delta(i,\tau)G(\tau,0), \nonumber\\
	\text{where} \quad R_0&\equiv&\frac{\det(I+B(\beta,\tau)(I+\Delta(i,\tau))B(\tau,0))}{\det(I+B(\beta,0))} \nonumber\\
	&=&1+\Delta_{i,i}(i,\tau)(I-G(\tau,\tau))_{i,i}.
	\label{eq:eq7}
\end{eqnarray}
Here we define $\Delta(i,\tau)=e^{V(s'(i,\tau))}e^{-V(s(i,\tau))}-I$ to update auxiliary field $s(i,\tau)$ and $B(\tau,\tau')=\prod_{\nu=\tau}^{\tau'}e^{H_s(\nu)}$ to represent product of decoupled partition function from imaginary time slices $\tau,\tau-1,...,\tau'$. In comparison, the nth Grover matrix can be written as
\begin{eqnarray}
	g_{A;s_1,s_2,...,s_n}&\equiv&\prod_i(G_{A;s_i})[I+\prod_j (G_{A;s_j}^{-1}-I)].
	\label{eq:eq8}
\end{eqnarray}

To simplify the notation, we further ignore the imaginary time label for Green's function $G(0,0)$ at zero imaginary time where reduced density matrices are defined accordingly. It's easy to generalize to zero-temperature version, where we just need to replace $G(\tau,0), G(0,\tau), G(0,0)$ with $G(\tau,\theta), G(\theta,\tau), G(\theta,\theta)$, where $\theta$ is the projection length towards the ground state and all the formulas and conclusions still hold~\cite{da2023teaching}. Update within $s_j$ will not affect $G_{A;s_i}$ for any $i\neq j$. Assuming we want to update within $s_1$, we define an auxiliary matrix
\begin{eqnarray}
	C_{A;s_1}&\equiv&G_{A;s_1}[I+\prod_j (G_{A;s_j}^{-1}-I)] \nonumber\\
	&=&I+G_{A;s_1}[I-\prod_{j>1} (G_{A;s_j}^{-1}-I)] \nonumber\\
	&\equiv&I+G_{A;s_1} M_{A;s_1}, \label{eq:eq9}
\end{eqnarray}
where $M_{A;s_1}\equiv I-\prod_{j>1} (G_{A;s_j}^{-1}-I)$ for notational convenience and remember $G_{A;s_1'}=G_{A;s_1}+G_{A;s_1}(0,\tau)\Delta(i,\tau)G_{A;s_1}(\tau,0)/R_0$. One will see $C_{A;s_1}$ serves as the Grover matrix in Ref.~\cite{d2022universal} and has a better stability when $n$ increasing. We then use the Sherman-Morrison formula to compute the updated $C_{A;s'_1}^{-1}$ and determinant ratio between two nth Grover matrices
\begin{eqnarray}
	&&C_{A;s'_1}^{-1}=C_{A;s_1}^{-1}(I+\frac{G_{A;s_1}(0,\tau)\Delta(i,\tau)G_{A;s_1}(\tau,0) M_{A;s_1}C_{A;s_1}^{-1}}{R_0 R_n}), \nonumber\\
	&&R_n\equiv\frac{\det(g_{A;s'_1,s_2,...,s_n})}{\det(g_{A;s_1,s_2,...,s_n})}=\frac{\det(C_{A;s_1}^{-1})}{\det(C_{A;s'_1}^{-1})} \nonumber\\
	&&= 1-\frac{1}{R_0}\Delta_{i,i}(i,\tau)G_{i,A;s_1}(\tau,0)M_{A;s_1}(C_{A;s_1})^{-1}G_{A,i;s_1}(0,\tau). \nonumber\\
	\label{eq:eq10}
\end{eqnarray}
In this formula, $R_0,G_{A;s_1}(\tau,0),G_{A;s_1}(0,\tau),\Delta(i,\tau)$ come from Green's function update routine and $M_{A;s_1}$ keeps the same when updating $s_1$ so that computing $R_n,C_{A;s'_1}^{-1}$ from $C_{A;s_1}^{-1}$ has the same complexity ($O(N^2)$) as the updating Green's function. Besides, one also needs formulae to update $s_i$ other than $i=1$ case and carry on numerical stablization. Similar with $B(\tau,\tau')$ matrix, we define $D(i,j)\equiv\prod_{k=i}^j(G_{A;s_{k}}^{-1}-I)$. Then we can write down $C_{A;s_i}^{-1}$ and $M_{A;s_i}$ in a simple form
\begin{eqnarray}
	C_{A;s_i}^{-1}&=&[I+D(i,n)D(1,i-1)]^{-1}G_{A;s_i}^{-1}, \nonumber\\
	M_{A;s_i}&=& I-D(i+1,n)D(1,i-1).
	\label{eq:eq11}
\end{eqnarray}
One can check the determinant of $C_{A;s_i}\prod_{j\neq i}(G_{A;s_j})$ is the same as $g_{A;s_1,s_2,...,s_n}$. And we use the formula above to carry on numerical stablization. To stably derive logarithm determinant of the nth Grover matrix, one should seperate the singular value matrix, trace the logarithm of it and add the logarithm determinant of other matrices.
\begin{figure}[htp!]
	\includegraphics[width=1\columnwidth]{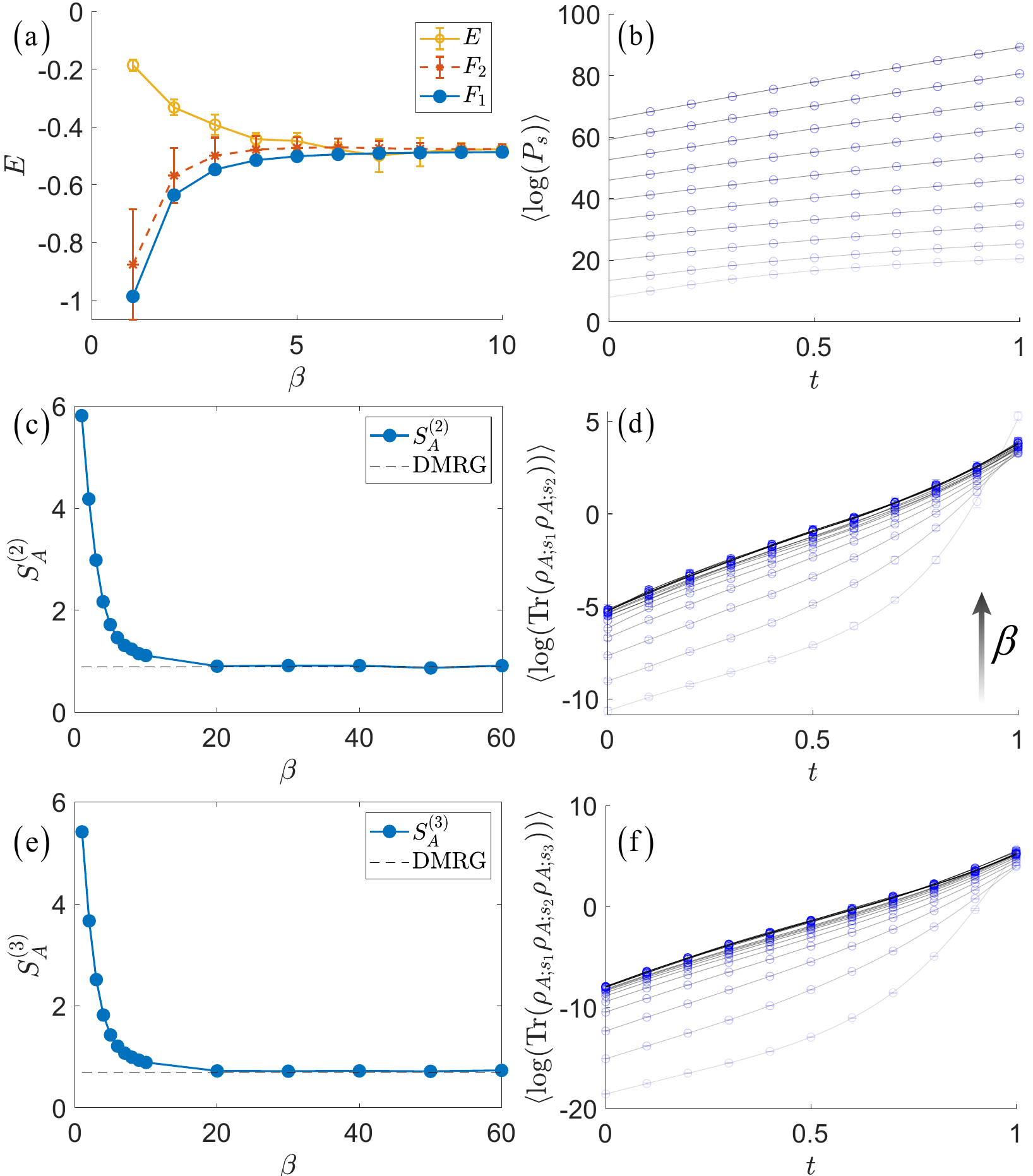}
	\caption{(a) Internal energy density $E$, free energy density $F_2$ derived by solving $E(\beta)=-\frac{\partial(\log(Z))}{\partial \beta}=\frac{\partial(\beta F_2)}{\partial \beta}$ and free energy density $F_1$ according to Eq.~\eqref{eq:eq6}. (b,d,f) $\left\langle \log(P_{s}) \right\rangle$, $\left\langle \log(\mathrm{Tr}(\rho_{A;s_1}\rho_{A;s_2})) \right\rangle$ and $\left\langle \log(\mathrm{Tr}(\rho_{A;s_1}\rho_{A;s_2}\rho_{A;s_3})) \right\rangle$ for different $\beta$ points used in (a,c,e) from small to large correspond to circles from bottom to top. Black lines show the fitting where the small slopes indicates small fluctuation for $\left\langle \log(P_{s}) \right\rangle$ at each $t$ point. (c) The 2nd EE $S_A^{(2)}$ converging with $\beta$ to 0.89, which is consistent with DMRG at zero temperature. (e) The 3rd EE $S_A^{(2)}$ converging with $\beta$ to 0.7, which is consistent with DMRG at zero temperature.}
	\label{fig:fig1}
\end{figure}

\noindent{\textcolor{blue}{\it Application to fermion Hubbard model.}---}
In this section, we present simulation results of free energy, the 2nd and 3rd R\'enyi EE for 2D square lattice fermion Hubbard model at half-filling
\begin{equation}
H=-t\sum_{\left\langle ij \right\rangle;s}c_{i;s}^{\dagger} c_{j;s}+U\sum_{i} (n_{i;\uparrow}+n_{i;\downarrow}-1)^2.
\label{eq:eq12}
\end{equation}
A few implementational considerations and the proof of absence of sign problem are listed in Appendix C, here we only focus on the results. In Fig.~\ref{fig:fig1}(a,b), we show the results for free energy for $L=4,U/t=8,\beta\in[1,10]$ with cylinder geometry. One can see free energy $F_1$ derived from Eq.~\eqref{eq:eq6} has a smaller fluctuation than internal energy. And the free energy $F_2$ derived from integrating internal energy data according to $E(\beta)=-\frac{\partial(\log(Z))}{\partial \beta}=\frac{\partial(\beta F_2)}{\partial \beta}$ even has a much larger error bar coming from $E$.
 
In Fig.~\ref{fig:fig1}(c,d), we show the results of the 2nd R\'enyi EE at the same parameter setting with entanglement area of $L\times L/2$ bi-partitioning a cylinder geometry to compare with DMRG result at zero temperature. The 3rd R\'enyi EE with the same parameter and the comparison with DMRG are shown in Fig.~\ref{fig:fig1}(e,f). Our results show a very small sampling error for both free energy and EE. The low temperature limit also matches well with DMRG.
\begin{figure}[htp!]
	\includegraphics[width=1\columnwidth]{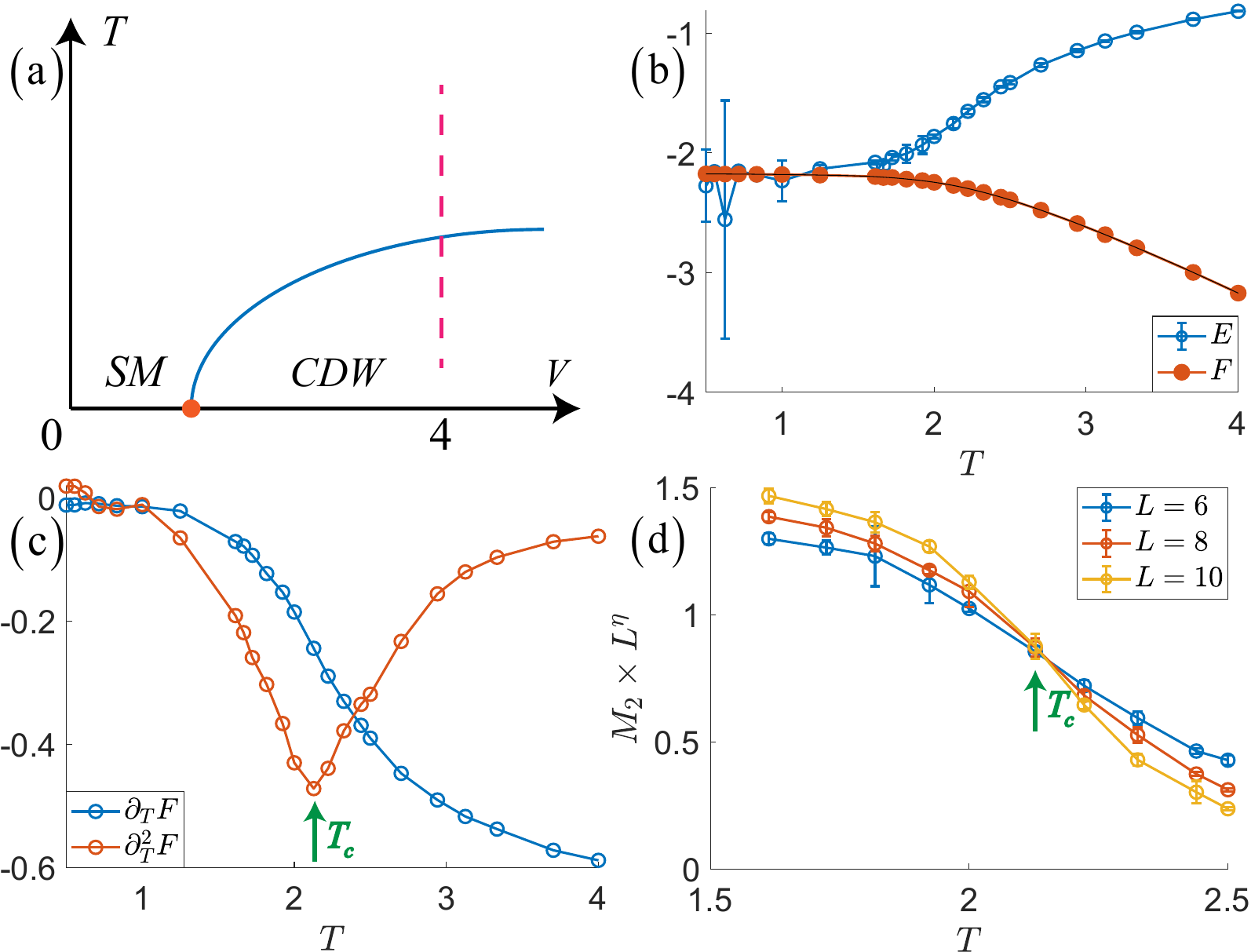}
	\caption{(a) Schematic phase diagram of spinless fermion $\pi$-flux half-filled square lattice t-V model. The orange point indicates N = 2 chiral-Ising transition in (2+1)D seperating Dirac semimetal ($SM$) and charge density wave ($CDW$) phases. The blue line indicates 2D Ising transition and our simulation goes along the dashed red line at $V=4$ where a finite temperature phase transition happens. (b) Directly measured internal energy density $E$ and free energy density $F$ measured by integral algorithm versus temperature $T$ for system size $L\times 2L$ with $L=6$ ($L=8,10$ have the similar internal energy and free energy density). (c) First-order ($\partial_T F$) and second-order derivative ($\partial_T^2 F$) of free energy for system size $L=6$, where the sharp dip with green arrow at $T_c\approx2.1$ in $\partial_T^2 F$ indicates the second-order phase transition temperature. (d) Finite temperature phase transition shown by green arrow at $T_c\approx2.1$ determined by 2D Ising finite size scaling ($M_2\equiv\sum_{i,j}\frac{\alpha_i\alpha_j}{L^4}\left\langle (n_i-\frac{1}{2})(n_j-\frac{1}{2}) \right\rangle$, where $\alpha_i=\pm1$ for $i\in A,B$ sublattice and $\eta=\frac{1}{4}$ for 2D Ising university class).}
	\label{fig:fig2}
\end{figure}

\noindent{\textcolor{blue}{\it Application to t-V model.}---}
In this section, we present simulation results of free energy for 2D square lattice spinless fermion $\pi$-flux t-V model at half-filling
\begin{equation}
	H=-\sum_{\left\langle ij \right\rangle}te^{i\phi_{ij}}c_{i}^{\dagger} c_{j}+V\sum_{\left\langle ij \right\rangle} (n_{i}-\frac{1}{2})(n_{j}-\frac{1}{2}).
	\label{eq:eq14}
\end{equation}
Here we set t=1 and take the gauge choice $e^{i\phi_{ij}}=1$ for $x$ direction $ij$ bonds and $e^{i\phi_{ij}}=\pm1$ for even(odd) column $y$ direction $ij$ bonds. The sign problem is proved to be free~\cite{Li2015,Wei2016,Wang2014} and the ground state transition from the Dirac semimetal to the charge-density-wave (CDW) insulator is found to belong to the Gross-Neveu chiral-Ising universality~\cite{Wang_Quantum_2023,Erramilli-Gross-2023}. We focus on large V (V=4) case for detecting the finite temperature 2D Ising phase transition from disorder phase to CDW phase from free energy. The result is shown in Fig.~\ref{fig:fig2}. One can see the free energy measurement is even better than internal energy as shown in Fig.~\ref{fig:fig2}(b) and the phase transition point indicated by second order derivative of free energy for a single size (Fig.~\ref{fig:fig2}(c)) matches well with the canonical finite size scaling result as shown in Fig.~\ref{fig:fig2}(d).

\noindent{\textcolor{blue}{\it Discussions.}---} 
In this Letter, we develop a fast update integral algorithm in DQMC, which converts measurements of exponential observables into conventional observables. This offers an elegant solution to the challenging problem of calculating exponential observables by fully suppressing the exponential fluctuations. 
Considering the current strong interests in entanglement entropy and determining the rich phase diagram of strongly-correlated spin/electronic systems, this highly efficient approach of computing R\'enyi entropy and free energy potentially has broad applications to a wide range of physical systems. In our illustration of 2D Hubbard model, the computed free energy has a smaller error than internal energy at large Hubbard U. In the example of t-V model, the derivative of free energy offers precise determinations of the phase transition. For systems with sophisticated interaction, such as twisted bilayer graphene and quantum Moir\'e systems~\cite{zhangMomentum2021,zhang2021superconductivity,panThermodynamic2022,huangEvolution2024}, our approach also offers much simpler way to access the free energy only from Green's function without invoking the Wick decomposition for the four fermion interactions. 
Besides, our method offers generic and easy access to the nth R\'enyi EE, negativity~\cite{wangEntanglement2023,wangEntanglementMicro2024} and entanglement spectrum~\cite{yanUnlocking2023,maoSampling2023} without incurring increases in computational complexity. 
Finally, our algorithm can also be generalized directly to the zero-temperature version projector QMC.

\begin{acknowledgements}
{\noindent \it Acknowledgements.---} XZ thanks Guangyue Han for inspiring information theory lectures. We thank Y. D. Liao for discussion on the integral algorithm. We thank the support from the Research Grants Council (RGC) of Hong Kong Special Administrative Region (SAR) of China (Projects Nos. 17301721, AoE/P701/20, 17309822,
C7037-22GF, 17302223), the ANR/RGC Joint Research Scheme sponsored by the RGC of Hong Kong SAR of China and French National Research Agency (Project No. A\_HKU703/22) and the HKU Seed Funding for Strategic Interdisciplinary Research. We thank HPC2021 system under the Information Technology Services, University of Hong Kong, and the Beijng PARATERA Tech CO.,Ltd. (URL:https://cloud.paratera.com) for providing HPC resources that have contributed to the research results reported within this paper.
\end{acknowledgements}

\appendix
\section{Implementational considerations for the integral algorithm}
We list a few technical considerations when performing a practical calculation. The first one is when we discretize the integral over $t$ naively (e.g., use $-\frac{1}{N_k}\sum_{i=1}^{N_k}\left\langle \log(\Tr(\rho_{A;s_1}\rho_{A;s_2}))\right\rangle_i$ to compute the 2nd R\'enyi entropy), the results must be larger than their true values. This indicates the function $\frac{\partial \log( f(t))}{\partial t}=\left\langle \log(\Tr(\rho_{A;s_1}\rho_{A;s_2}))\right\rangle_t$ is an increasing function. We notice the second-order derivative $\frac{\partial^2 \log( f(t))}{\partial t^2}=\left\langle \log(\Tr(\rho_{A;s_1}\rho_{A;s_2}))^2\right\rangle_t-\left\langle \log(\Tr(\rho_{A;s_1}\rho_{A;s_2}))\right\rangle_t^2\geqslant0$ indicates the fluctuation for $\left\langle \log(\Tr(\rho_{A;s_1}\rho_{A;s_2}))\right\rangle_t$. As long as the slope is not sharp (i.e., indicating mild fluctuation), we can compute $\left\langle \log(\Tr(\rho_{A;s_1}\rho_{A;s_2}))\right\rangle_t$ accurately at a given sequence of $t$ points and numerically integrate within $[0,1]$ to derive the 2nd R\'enyi entropy. The corresponding results are shown in main text Fig. 1 (c,d).

The second point is that one may observe the sampling may not be "importance sampling" for a small $t$ when computing free energy. One can check the formula for free energy at $t\rightarrow0$ limit, where the weight becomes independent with determinant $P_s$ as well as the Hamiltonian so that one actually updates auxiliary fields randomly. This can also be seen with the increasing acceptance ratio as $t\rightarrow0$. One way to avoid this inefficient sampling is to update by propose-accept/reject method for small $t$. Assuming $n\approx1/t$, we can propose $n$ auxiliary fields update and then determine accepting it or not. This will recover the normal acceptance ratio as $t=1$.

The third point is about the computation of $\log(\Tr(\rho_{A;s_1}\rho_{A;s_2}))$ or $\log(P_s)$ for each configuration. Directly computing the determinant will derive exponentially small (or large) number and the exponent is hard to compute accurately by $\log$. A more efficient way is to utilize the numerical stabilization step for computing, say logarithm determinant of Green's function.
\begin{widetext}
\begin{eqnarray}
G(\tau,\tau)&\equiv&(I+B(\tau,0)B(\beta,\tau))^{-1} \nonumber\\
&\equiv&(I+U_RD_R^{>}D_R^{<}V_RV_LD_L^{<}D_L^{>}U_L)^{-1} \nonumber\\
&=&U_L^{-1}(D_L^{>})^{-1}[(D_R^{>})^{-1}(U_LU_R)^{-1}(D_L^{>})^{-1}+D_R^{<}V_RV_LD_L^{<}]^{-1}(D_R^{>})^{-1}U_R^{-1} \nonumber\\
\log(\det(G(\tau,\tau)))&=&-\log(\det([(D_R^{>})^{-1}(U_LU_R)^{-1}(D_L^{>})^{-1}+D_R^{<}V_RV_LD_L^{<}])) \nonumber\\
&-&\log(\det(U_L))-\log(\det(U_R))-\Tr(\log(D_L^{>}))-\Tr(\log(D_R^{>}))
\end{eqnarray}
\end{widetext}
Here for diagonal positive matrix $D$, we decompose it according to the diagonal element larger ($D^{>}$) or smaller ($D^{<}$) than 1, i.e. $D\equiv D^{>}D^{<}$. In this way, one can compute $\log(\det(G(\tau,\tau)))$ stably for each configuration sample.

\section{Derive integral formula by taking limit of the incremental algorithm}
The same formula Eq. (2) can also be derived from taking small polynomial power limit of the incremental algorithm~\cite{da2023controllable,da2023teaching,pan2023stable}. Below, we will illustrate those steps. In incremental algorithm, $S_{A}^{(2)}$ is computed as
\begin{eqnarray}
S_{A}^{(2)}&=&-\sum_{i=1}^{N_k}\log(\frac{\lambda_i}{\lambda_{i-1}}), \nonumber\\
\frac{\lambda_i}{\lambda_{i-1}}&=&\frac{\sum P_{s_1}P_{s_2}\Tr(\rho_{A;s_1}\rho_{A;s_2})^{(i-1)/N_k} \Tr(\rho_{A;s_1}\rho_{A;s_2})^{1/N_k}}{\sum P_{s_1}P_{s_2}\Tr(\rho_{A;s_1}\rho_{A;s_2})^{(i-1)/N_k}} \nonumber\\
&\equiv&\left\langle \Tr(\rho_{A;s_1}\rho_{A;s_2})^{1/N_k}\right\rangle_i. 
\label{eq:eq4}
\end{eqnarray}
The computation for each piece $-\log(\left\langle \Tr(\rho_{A;s_1}\rho_{A;s_2})^{1/N_k}\right\rangle_i)$ can be computed in parallel and adding them together will derive the final result in Ref.~\cite{da2023controllable}. We notice the distribution $x^{1/N_k}$ for random variable $x>0$ will approach to uniform distribution when taking $N_k\rightarrow\infty$. Using Jensen's inequality then take the equality case when we set $N_k\rightarrow\infty$ limit as

\begin{eqnarray}
S_{A}^{(2)}&=&-\sum_{i=1}^{N_k}\log(\left\langle \Tr(\rho_{A;s_1}\rho_{A;s_2})^{1/N_k}\right\rangle_i) \nonumber\\
&\leqslant&-\frac{1}{N_k}\sum_{i=1}^{N_k}\left\langle \log(\Tr(\rho_{A;s_1}\rho_{A;s_2}))\right\rangle_i \nonumber\\
&\rightarrow&-\int_{0}^{1} dt \frac{\sum P_{s_1}P_{s_2}\Tr(\rho_{A;s_1}\rho_{A;s_2})^{t} \log(\Tr(\rho_{A;s_1}\rho_{A;s_2}))}{\sum P_{s_1}P_{s_2}\Tr(\rho_{A;s_1}\rho_{A;s_2})^{t}} \nonumber\\
&\equiv&-\left\langle \log(\Tr(\rho_{A;s_1}\rho_{A;s_2}))\right\rangle_{s_1,s_2;t}.
\label{eq:eq5}
\end{eqnarray}
Since we take $N_k\rightarrow\infty$ limit for integral, $S_{A}^{(2)}=-\left\langle \log(\Tr(\rho_{A;s_1}\rho_{A;s_2}))\right\rangle_{s_1,s_2;t}$ is exact. Here, our observable becomes $\log(\Tr(\rho_{A;s_1}\rho_{A;s_2}))$ which is not exponentially small and the integral over continuous field $t\in[0,1]$ can be realized by numerical integral. This just gives the same formula as Eq. (3)

\section{Proof of sign problem free in Hubbard model}
We would like to prove there is no sign problem when computing the exponential observables in Hubbard model. Because of the well known particle-hole transformation in density decoupling channel for bipartite lattices, we define $\tilde{c}_{i;\downarrow}^{\dagger}\equiv (-1)^{i}c_{i;\downarrow}$ and have $(I+\tilde{B}_{\downarrow})_{i,j}^{-1}=(I+B_{\uparrow}^*)_{i,j}^{-1}$ so that the weight $P_s\equiv\det(I+B_{\uparrow})\det(I+B_{\downarrow})$ is always non-negative. This indicates directly the simulation for free energy should be sign-problem-free since the weight there is just $P_s^t$. As for the proof for the nth R\'enyi EE, we need to prove the determinant of Grover matrix is non-negative for any auxiliary field configuration. First we need a relation $(-1)^{i+j}(I-G_{j,i;\downarrow})=\tilde{G}_{i,j;\downarrow}\equiv(I+\tilde{B}_{\downarrow})_{i,j}^{-1}=(I+B_{\uparrow}^*)_{i,j}^{-1}=G_{i,j;\uparrow}^*=G_{i,j;\downarrow}^*$, so that $U_c^{-1}(I-G_{\downarrow})U_c=G_{\downarrow}^{\dagger}$ where $U_c$ is the constant unitary transformation giving $(-1)^{i}$ coefficient according to sublattice label. Then we have
\begin{eqnarray}
&&U_c^{-1}D_{s_i} U_c=U_c^{-1}(G_{A;s_i}^{-1}-I) U_c=(D_{s_i}^{\dagger})^{-1}, \nonumber\\
&&U_c^{-1}(\prod_{i}D_{s_i})U_c=((\prod_{i}D_{s_i})^{\dagger})^{-1}. \label{eq:eq13}
\end{eqnarray}
For convenience, we denote $e^{\Gamma_{\alpha}}$ as the eigenvalue for $\prod_{i}D_{s_i}$ and $e^{\lambda_{\alpha;s_i}}$ as the eigenvalue for $D_{s_i}$. According to Eq.~\eqref{eq:eq13} we have $e^{-\Gamma_{\alpha}^*}=e^{\Gamma_{\beta}}$ and $e^{-\lambda_{\alpha;s_i}^*}=e^{\lambda_{\beta;s_i}}$ where $\alpha$ and $\beta$ can label the same or different eigenvalues and form an injective function. Besides, we have $\det(\prod_{i}D_{s_i})=e^{\sum_{\alpha}\Gamma_{\alpha}}=e^{\sum_{i,\alpha_i}\lambda_{\alpha_i;s_i}}$. Now we are ready to compute the determinant of Grover matrix which is $\det(g_{A,s_1,...,s_n})=\prod_{i}\det((I+D_{s_i})^{-1})\det(I+\prod_{i}D_{s_i})$ according to Eq.~\eqref{eq:eq8}.
\begin{eqnarray}
g_{A;s_1,s_2,...,s_n}&\equiv&\prod_i(G_{A;s_i})[I+\prod_j (G_{A;s_j}^{-1}-I)].
\label{eq:eq8}
\end{eqnarray}
There are many terms summation after writing in $\prod_{i}D_{s_i}$ and $D_{s_i}$ diagonal basis (i.e., $\det(g_{A,s_1,...,s_n})=\sum_{M}\frac{e^{\sum_{\alpha\in M}\Gamma_{\alpha}}}{\sum_{m_i}e^{\sum_{i,\alpha_i\in m_i}\lambda_{\alpha_i;s_i}}}$ where $M$ and $m_i$ label all the possible choices of choosing eigenvalue sets from $\{\Gamma\}$ and $\{\lambda_{s_i}\}$) . For any term written as $\frac{e^{\sum_{\alpha\in M}\Gamma_{\alpha}}}{\sum_{m_i}e^{\sum_{i,\alpha_i\in m_i}\lambda_{\alpha_i;s_i}}}$, the complex conjugate of it is also in this summation, i.e.
$(\frac{e^{\sum_{\alpha\in M}\Gamma_{\alpha}}}{\sum_{m_i}e^{\sum_{i,\alpha_i\in m_i}\lambda_{\alpha_i;s_i}}})^*
=\frac{e^{\sum_{\alpha\notin M'}\Gamma_{\alpha}}}{\sum_{m_i}e^{\sum_{i,\alpha_i\notin m_i'}\lambda_{\alpha_i;s_i}}}$ where $M',m_i'$ are mapped eigenvalue set from $M,m_i$. Since the maps from $M,m_i$ to $M',m_i'$ are injective, for any given $M$ and term in the summation, one can always find one and only one which is complex conjugate of it so that $\det(g_{A,s_1,...,s_n})$ is always real for one spin sector and non-negative after square.

\bibliographystyle{apsrev4-2}
\bibliography{EE_QMC}

\end{document}